 \documentclass[times]{qjrms4}

\usepackage{soul}

\usepackage[colorlinks,bookmarksopen,bookmarksnumbered,citecolor=blue,urlcolor=blue]{hyperref}

\newcommand\BibTeX{{\rmfamily B\kern-.05em \textsc{i\kern-.025em b}\kern-.08em
T\kern-.1667em\lower.7ex\hbox{E}\kern-.125emX}}

\usepackage{moreverb}

\usepackage{natbib}
\def\volumeyear{2016}

\begin{document}

\runningheads{D.~Paz\'o \emph{et~al.}}{Data-assimilation by delay-coordinate nudging}

\title{Data-assimilation by delay-coordinate nudging}

\author{D.~Paz\'o,\affil{a}\corrauth  A. Carrassi\affil{b} and J.~M.~L\'opez\affil{a}}

\address{\affilnum{a} Instituto de F{\'\i}sica de Cantabria (IFCA), 
 CSIC-Universidad de Cantabria, Santander, Spain \\
\affilnum{b} Nansen Environmental and Remote Sensing Center, Bergen, Norway}

\corraddr{Diego Paz\'o, Instituto de F{\'\i}sica de Cantabria, Avenida Los Castros, 39005 Santander, Spain}

\begin{abstract}

A new nudging method for data assimilation, {\em delay-coordinate nudging}, is presented. 
Delay-coordinate nudging makes explicit use of present and past observations in
the formulation of the forcing driving the model evolution at each time-step. 
Numerical experiments with a low order chaotic system show that the
new method systematically outperforms standard nudging in different model and observational
scenarios, also when using an un-optimized formulation of the delay-nudging coefficients. 
A connection between the optimal delay and the dominant Lyapunov exponent
of the dynamics is found based on heuristic arguments and is confirmed by the
numerical results, providing a guideline for the practical implementation of the algorithm. 
Delay-coordinate nudging preserves the easiness of implementation, the intuitive
functioning and the reduced computational cost of the standard nudging,
making it a potential alternative especially in the field of seasonal-to-decadal
predictions with large Earth system models that limit the use of more sophisticated 
data assimilation procedures.

\end{abstract}


\maketitle

\section{Introduction}

State estimation and optimal control theory in geoscience is commonly 
referred to as data assimilation (DA) \citep{Jazwinski}. The term encompasses 
the entire sequence of operations that, starting from the observations 
of a system, and possibly from additional statistical and/or dynamical 
information (e.g.~a model), provides the best-possible estimate 
of its state \citep{Kalnay}.

DA methods are usually grouped into variational and sequential. 
In four-dimensional variational (4DVar) DA the model 
trajectory is adjusted to fit the observations distributed within a given 
time interval \citep{Sasaki1970}. The 4DVar, with some specific approximations 
aimed at reducing its computational cost, has been successfully applied to 
atmospheric and oceanic models and it is in particular adopted at the European 
Centre for Medium range Weather Forecast, ECMWF, \citep{Rabier2000}, at 
M\'et\'eo-France and at the UK MetOffice. In parallel to the variational approach, 
ensemble-based methods represent a successful and fruitful alternative. In this 
case, the DA problem is formulated in a sequential way with a forecast step, 
where the state estimate and its associated error statistic are propagated in time, 
alternated with the analyses steps when these estimates are updated using the 
observations. The ensemble-based schemes are Kalman-filter-like algorithms 
\citep{Kalman1960} in which the error description is obtained using an ensemble of 
model trajectories aimed at representing the first two  moments of the unknown error 
probability density function. The best-known ensemble-based 
scheme is the ensemble Kalman filter (EnKF; \cite{Evensen}); a version of the 
EnKF is nowadays operational for the atmospheric model at the Canadian Meteorological 
Centre, CMC \citep{Houtekamer2014} and for the ocean system by MET Norway 
\citep{Sakov2012}.
While all these DA methods are formulated in a Gaussian framework, and have 
proven
to work quite well with large-dimensional systems, a new stream of research 
is nowadays considering the introduction of fully Bayesian procedures 
(see e.g.~\citep{Bocquet2010}).  

Among all sequential DA methods, nudging is probably the simplest one.
It is an empirical DA technique, 
introduced in the early '70 (see e.g.~\cite{Anthes74, hoke76,nudging}).
The approach, inspired by control system  
theory \citep{Gelb},  
consists in adding a term to the prognostic equations that, acting like an 
extra-coupling term, drives the model trajectory toward 
the observations of the unknown system intended to be estimated. 
The coupling strength is expressed as a relaxation 
timescale \citep{Macpherson91}, and is usually chosen on the basis of the 
properties of the variable to be nudged.

The empirical origin of nudging has not precluded its formalization 
and the outline of connections with synchronization problems 
and Kalman filtering \citep{duane06,szendro_jgr}.
Methods to estimate systematically the optimal nudging coefficients
were proposed by \cite{Zou92,stauffer93}. This was followed
by more sophisticated approaches where the coefficients are made 
flow-dependent, developed in parallel
by meteorological \citep{vidard03,auroux08,auroux09}
and dynamical systems \citep{so94,junge01} communities.

A well-known technique of time series analysis consists in the
use of delayed coordinates, since this permits the attractor reconstruction
from a limited number of observed variables (see e.g.~\citep{KS}). 
The idea of ``temporal embedding'', put forward by \cite{packard80}
and mathematically proven by \cite{takens81}, has been extraordinary
fruitful; and has been recently incorporated into synchronization
studies \citep{abarbanel09} (see also \citep{rey_pla,rey_pre}). 
This approach, putting aside model errors and observational noise, 
is based on the expectation
that coupling the model to the current and past observations
with certain time-dependent nudging coefficients permits full synchronization 
between model and reality.
However, finding the time-dependent nudging coefficients
is a hard mathematical task \citep{rey_pla,rey_pre}, with difficult implementation
in a realistic model (and certain theoretical limitations \citep{parlitz14}).
{\color{black} One example of such difficulties and proposed solution is given in \citep{rey_pla}, 
where} the Jacobian of the map from physical to 
the time-delay space is used to compute the forcing terms.

As an alternative, in this paper we investigate a simple albeit effective way of improving
the performance of classical nudging, in which nudging coefficients are static
and assimilation is performed in a sequential way (i.e.~``on the fly'').
The new method, named {\it delay-coordinate nudging}, makes explicit use of 
past observations 
incorporating them to
the forcing terms at each time step of the model evolution.
The new method systematically outperforms standard nudging
(also in the presence of model error), while
the increase in computational cost with respect to classical nudging 
(without delay coordinates) is negligible. Our implementation preserves 
the easiness of implementation, the intuitive functioning and the 
reduced computational cost, that have made nudging the preferred choice 
in the growing field of initialized seasonal-to-decadal predictions 
with coupled Earth system simulators \citep{Magnusson2013,carrassi14,meehl13,doblas13,sanchez-gomez,servonnat}. 

Obviously, the number 
of parameters to be optimized in delay-coordinate nudging is larger than in standard nudging, 
as the coupling strength at each 
observation within the delay interval has to be determined, 
along with the length of the interval itself.   
However, our numerical results with a low-order nonlinear dynamical system show 
that by setting the coupling strengths equal at each past observation 
only slightly deteriorates the performance with
respect to the fully 
optimized case, but with a substantial reduction in the computational cost.
Furthermore it is shown that the optimal delay is related to the dominant Lyapunov 
exponent of the attractor. This provides a physical rationale in support of 
the delay-coordinates nudging, and a guideline to choose the optimal delay. 

The paper is structured as follows: classical and delay-coordinate nudging are 
described in sections 2 and 3, respectively; numerical results with
both perfect and 
imperfect model setups are reported in sections 4 and 5. Section
6 presents  
a mathematical argument for
the delay-coordinate nudging 
that is intended to clarify the theoretical basis for the new
method. Final conclusions are drawn in section 7. 

\section{Classical Nudging}
We begin by reviewing standard nudging.
To simplify the presentation let us consider a
spatially discretized system defined in a one-dimensional lattice
(e.g.~on a latitude circle) and
assume that the state of the system
at each position is given by a scalar quantity.
Furthermore, let ${\mathbf u}=(u_1,u_2,\ldots,u_N)$ be the $N$-dimensional state
vector 
representing the unknown system we intend to estimate by doing DA.
We assume the `truth' evolves continuously in time and its dynamics be described
by the 
following autonomous dynamical system:
\begin{equation}
\dot{\mathbf{u}} = {\mathbf{g}}({\mathbf{u}}),
\end{equation}
where upper dot represents the time derivative 
and ${\mathbf{g}}$ is the vector field.
In most applications with practical relevance, only a portion of the system's state vector 
is observed, so that observations are performed only at a subset $\cal{M}$ 
of the $N$ sites; $dim({\cal M})=M<N$. 
The second key ingredient in DA is the model; 
it summarizes the knowledge about the unknown dynamics and it is written here 
under the form of the autonomous dynamical system     
 $\dot{\mathbf{v}} =\tilde{\mathbf{g}}({\mathbf{v}})$,
where $\mathbf{v}$ and and $\tilde{\mathbf{g}}$ are the state vector 
and the vector field of the model, respectively. In the ideal case
of a perfect model $\mathbf{v}$ is $N$-dimensional and $\tilde{\mathbf{g}}=\mathbf{g}$.

DA is performed to optimally use the observations and the model
in order to ``reconstruct'' the whole signal, ${\mathbf{u}}$. This
accounts to  
make use of the information contained in the observed areas and
gain knowledge on the unobserved ones.
 
Nudging can be seen as one of the simplest ways to achieve this goal, in a
hierarchy of 
DA procedures of increasing complexity and computational cost
\citep{LesHouches2014}.

For the sake of focusing on the problem of the spatial sparseness of the
observations, we assume hereafter a continuous access to error-free observations, so there is no
need for smoothing (i.e.~interpolate in time) the observational data.

The classical nudging algorithm consists in introducing 
a dissipative coupling between the truth and the model (`master' and `slave' in the
dynamical systems' jargon):
\begin{equation}
\dot {\mathbf{v}} =\tilde{\mathbf{g}}({\mathbf{v}}) + \kappa \mathbf{D}  ({\mathbf{u}} - {\mathbf{v}} )
\label{nudging}
\end{equation}
where $\mathbf{D}$ is a diagonal matrix with elements $D_{ii}=1$ 
if the $i$-th site is observed, and $D_{ii}=0$ otherwise.

The scalar coefficient $\kappa$ controls the coupling strength
and is chosen here to be site-independent
because, for the sake of simplicity, 
the system is assumed to be homogeneous in space and the
observed sites are evenly spaced. This situation is typical 
 when running hindcast simulations with an Earth system model nudged toward a 
 reanalysis dataset which gives climate variables on a regular spatial and temporal interval. 

The testbed for our discussion and comparison is
the Lorenz-96 model \citep{lorenz96}. This classical model is an atmospheric
low-order representation of the dynamics of a scalar variable on a circle at
equal latitude in the extra-tropics.
We shall consider first the case in which the model geometry is 
a single ring and there is only one temporal scale.
The system of $N$ ordinary differential equations governing the truth reads:
\begin{equation}
\dot{u}_i = u_{i-1}\left(u_{i+1} -u_{i-2}\right)-u_i+F \quad i=1,...,N, \label{l96m}
\end{equation}
where $i$ runs from $1$ to $N$ with periodic boundary conditions: $u_0=u_N$, $u_{-1}=u_{N-1}$, $u_{N+1}=u_1$.
Here, we adopt a system size of $N=60$, and the parameter $F=8$ (unless stated otherwise).
With these choices the model behaves with well-developed spatio-temporal chaos \citep{pazo08}. 
The chaotic attractor has
$20$ positive Lyapunov exponents, being $\mu \simeq 1.75$ the largest one.
The numerical integration of Eq.~\eqref{l96m} is performed with
an explicit Euler scheme with time step $dt=10^{-3}$, which is hence also the assimilation frequency.
We anticipate that, in Sec.~\ref{imperf}, we shall investigate the effect of
adopting another 
version of the Lorenz-96 model with two time scales.

In the nudging method the model is coupled to the truth according to the
general scheme in Eq.~\eqref{nudging}. In particular, for a truth given by Eq.~\eqref{l96m}
we have (in a perfect model scenario)
\begin{equation}
\dot{v}_i = v_{i-1}\left(v_{i+1} -v_{i-2}\right)-v_i+F 
+\begin{cases}  
\kappa(u_i-v_i) & \mbox{if $i\in {\cal M}$}
\\
0  & \mbox{if $i\notin \cal{M}$}
\end{cases}   \label{l96s}
\end{equation}
where $\cal M$ is the subset of $M$ observed sites.
For instance, if the observations are  available every $n_s\equiv N/M=2$ sites then
we take ${\cal M}=\{1,3,5\ldots\}$ (taking the even sites would be equivalent since the system
is homogeneous and has periodic boundary conditions).
In this study $n_s$ is mainly equal to $3$ or $4$. Given the highly chaotic nature of the 
dynamics (note that the number of positive Lyapunov exponents is no less than
$1/3$ of the model size) 
the state estimation with $n_s=3$ ($M=20$) or $n_s=4$ ($M=15$) is
already highly challenging since  
observations have a sizeable sparsenesses as compared with the total number of
unstable directions.

\begin{figure}
\centering
\includegraphics[width=19pc]{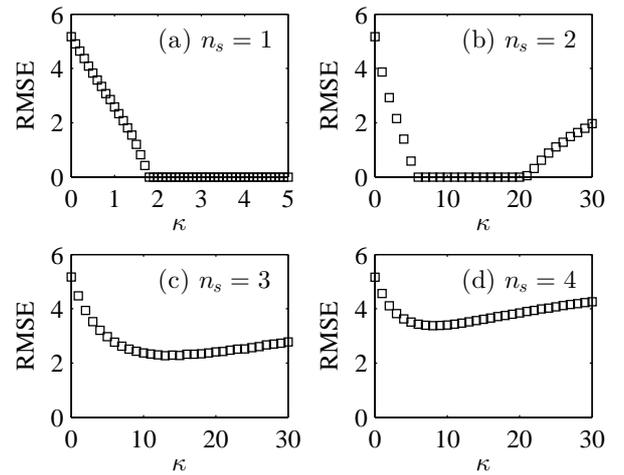}
\caption{RMSE for different degrees of coupling sparseness $n_s=1-4$ from (a) to (d).
For $n_s=1$ (dense coupling) the RMSE vanishes for $\kappa>\mu\approx1.75$. 
For $n_s=2$, the RMSE is zero in a certain range of $\kappa$.
Finally, for $n_s\ge3$ the observations are too sparse and the RMSE remains positive for all $\kappa$,
with an optimal coupling at $\kappa\approx13$ and $\kappa\approx8$, for $n_s=3$ and $n_s=4$, respectively.
{\color{black} For each data point, the model was initialized with a random error of order 0.1 at each site.}
}
\label{nodelay}
\end{figure}

To quantify the performance of the nudged model to describe the truth we compute
the root-mean-square error:
\begin{equation}
 {\mathrm{RMSE}}=\left< \left[ \frac{1}{N} \sum_{i=1}^N (v_i-u_i)^2 \right]^{1/2} \right>
\end{equation}
where the brackets denote a temporal average, which we take over one long realization 
of typically  5$\times 10^4$ t.u.,  
{\color{black} after a transient of 5$\times 10^2$ t.u.
We made sure both that our results are independent of the initial conditions,
and that our simulations were run long enough for the truth-model system to reach the stationary state.}
Figure \ref{nodelay} shows the results of our
numerical
simulations for different observational sparsenesses:  $n_s=1,2,3$ and $4$.
For dense coupling ($n_s=1$), the RMSE converges to zero for couplings
$\kappa$ above a critical value $\kappa_c$,
which equals the largest Lyapunov exponent of the model, $\kappa_c=\mu\approx 1.75$,
in this case. 
For $n_s=2$ the RMSE drops to zero at a critical value of $\kappa$, but becomes nonzero
again when the coupling becomes too large (due to a kind of overfitting effect
as the coupling to observations increases). For $n_s\ge3$, the observations are
too sparse and the
RMSE remains nonzero for any coupling strength. This is related 
to the high `chaoticity' of the Lorenz-96 model, with $1/3$ of unstable directions,
for the parameters used in this study. However, despite the RMSE does not drop to
zero for $n_s \ge 3$,
there is an optimal coupling ($\kappa\approx13$ for $n_s=3$) for which the RMSE
reaches its global minimum (or `infimum'):
$\inf\limits_\kappa({\mathrm{RMSE}})\approx\mathrm{RMSE}(\kappa=13)\approx 2.28$.
Analogous effect appears for $n_s=4$ in Figure~\ref{nodelay}(d),
with an optimal $\kappa\approx8$ and $\inf\limits_\kappa({\mathrm{RMSE}})\approx3.37$.

Up to this point we have briefly discussed the effect of classical nudging for
DA of sparse observations in the Lorenz-96 model in a very demanding
high-complexity setup with up $1/3$ of all tangent-space directions being
unstable. Next, we introduce the new method of delay-coordinate nudging and
compare its performance with the classical nudging scheme just discussed.

\section{Delay-coordinate nudging}

We now propose a modification of the classical nudging scheme in
Eq.~(\ref{nudging})
to incorporate also dynamical information from delayed coordinates. The
new DA scheme takes the form
\begin{equation}
\dot{\mathbf{v}} = \tilde{\mathbf{g}}({\mathbf{v}}) + \left[\sum_{n=0}^{P-1} 
\kappa_n \mathbf{D} 
({\mathbf{u}}^{n\tau} - {\mathbf{v}}^{n\tau}) \right],
\label{delayed}
\end{equation}
where the superscript $n\tau$ denotes the time interval of delay, such that
$\mathbf{v}^{n\tau}\equiv\mathbf{v}(t-n\tau)$, and $P$ is the total number
of (present plus delayed) observations used at each observed site. 
Therefore, in contrast to Eq.~\eqref{nudging} the model is now driven not only by
the discrepancy with the observations at present time ($n=0$), but also by a
finite number of past observations ($n\ge1$), each of them contributing
with an independent coupling strength $\kappa_n$.

In the context of dynamical systems theory it is well known that, 
if some variables of a system cannot be measured, the attractor's topology
can be nonetheless reconstructed using delayed variables, according to 
the Takens's delay embedding theorem \citep{KS}.
This indicates that delayed variables convey useful information 
that only very recently has been considered for DA purposes
\citep{rey_pla,rey_pre}. 
It is worth noting that the use of delayed coordinates in Eq.~\eqref{delayed}
does not imply the need for 
additional observations, but just the storage and use of a finite number of
previous measurements. 
Also note that the delay-coordinate nudging scheme that we introduce in this
paper constitutes the simplest method of using the information from 
past observations and other, more involved, schemes can be envisaged.
The increase of computational cost of the new algorithm relative to the standard nudging is         
negligible, and the new procedure keeps the easiness of implementation that makes 
the nudging attractive for applications in large-dimensional systems. 

Further theoretical justification for the validity of the
new delay-coordinate nudging algorithm is postponed to Sec.~\ref{sec_theory}.

\begin{figure*}
\centering
\includegraphics{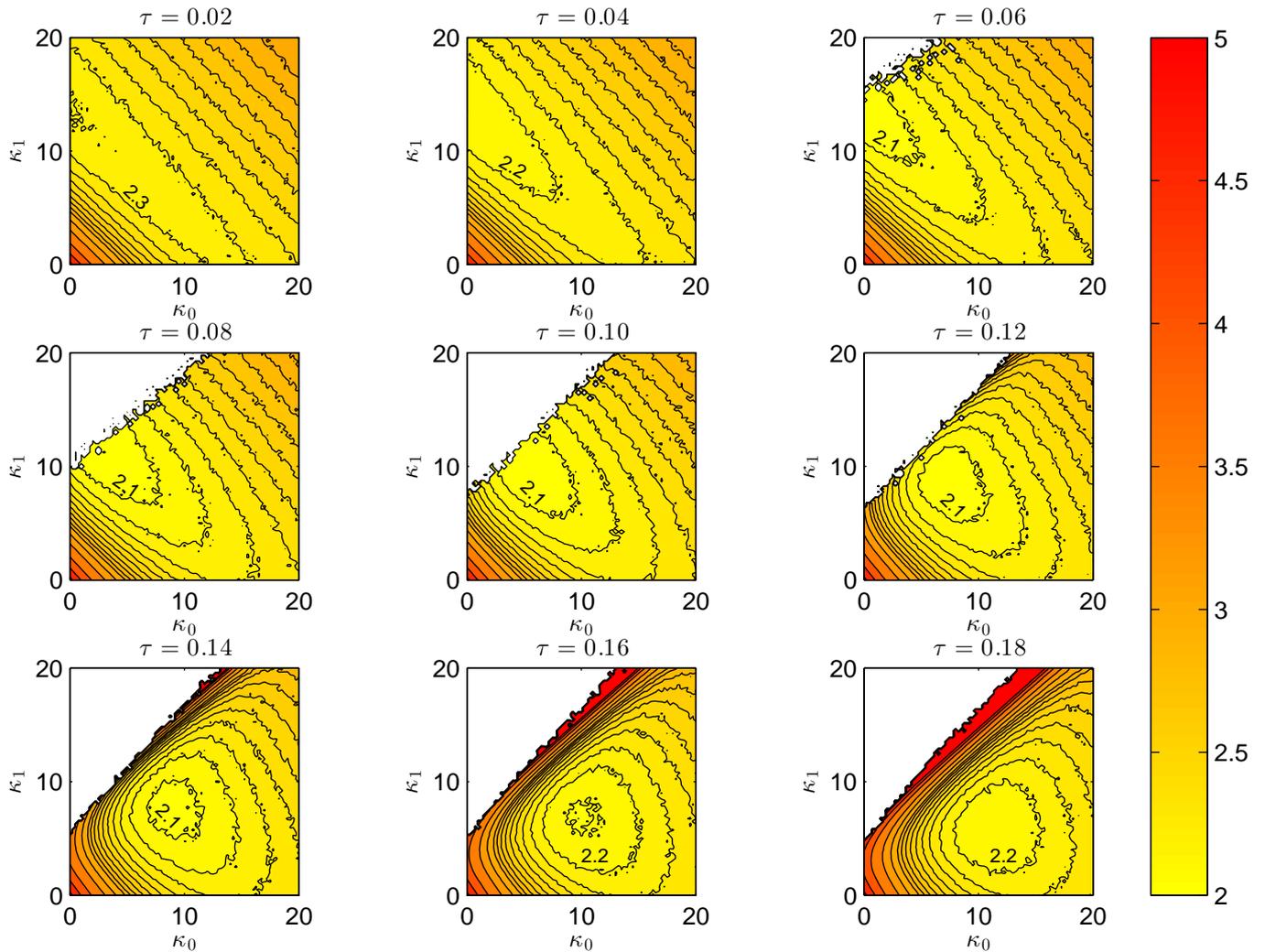}
\caption{Colour shading depicting the RMSE for $n_s=3$ and several delays.
In the most yellowish region the
RMSE is below $2.1$. In the small RMSE regions, the contour lines are at levels separated
by intervals of $0.1$ in RMSE. In several subplots there is a white region in the upper left corner
indicating where the algorithm is found to be unstable.}
\label{ns3}
\end{figure*}

\section{Numerical Analysis: Perfect model scenario}

We start considering a perfect model scenario, $\tilde{\mathbf{g}}=\mathbf{g}$, and
implement delayed nudging in the Lorenz-96 model using only 
one delayed coordinate at each observation site (i.e.~$P=2$).
The equation for the model, forced by the nudging terms, reads:
\begin{eqnarray}
\dot{v}_i &=& v_{i-1}\left(v_{i+1} -v_{i-2}\right)-v_i+F \nonumber \\
&+&\begin{cases}  
\kappa_0(u_i-v_i) + \kappa_1 (u_i^\tau-v_i^\tau)& \mbox{if $i\in {\cal M}$}
\\
0  & \mbox{if $i\notin \cal{M}$}
\end{cases}  
\label{l96d}
\end{eqnarray}

Given the results depicted in Figure~\ref{nodelay} for the standard nudging without delay, 
we concentrate the attention on studying the delayed coordinate approach when $n_s\ge3$.

\subsection{Case I: $n_s=3$}

In Figure~\ref{ns3} we present our results for a sparseness $n_s=3$, i.e.
when observations are available every $3$
sites, so that the subset of observed sites is ${\cal M}=\{1,4,7\ldots\}$.
For several values of $\tau\in[0.02,0.2]$, we have 
varied $\kappa_0$ and $\kappa_1$ with step-sizes $\Delta k=0.25$.

The standard nudging without delay is recovered in Figure~\ref{ns3} along the
$\kappa_0$-axis, 
since for $\kappa_1=0$ the delay is irrelevant, and the isolines intercept the 
axis at the same points in all panels.
The lowest RMSE, the clearest yellow in the figure, is in the range $2-2.1$; 
note that the minimal RMSE without delay is $2.28$ (Sec.~2).  
The lowest RMSE region appears for $\tau$ as large as $0.06$ and progressively 
moves away from the $\kappa_1$ axis
as $\tau$ increases, eventually disappearing for $\tau\ge0.18$.
This behaviour seems to suggest the existence of an optimal delay $\tau$. Also,
it shows that, 
for increasing $\tau$, a larger weight has to be given to the present time
observations 
(increasing $\kappa_0$ and decreasing $\kappa_1$).

While results show that delay allows to achieve smaller RMSE than in the classic
approach with no delay, 
they also reveal that caution has to be taken when choosing the forcing strength. In fact, 
delayed nudging becomes unstable in certain regions of the $\kappa_0-\kappa_1$ plane 
(see e.g.~$\tau = 0.06$ in Figure~\ref{ns3}). The region of divergence is displayed
in white in Figure~\ref{ns3}. In agreement with the theoretical argument
outlined in Sec.~\ref{sec_theory}, 
those white regions remain above the bisector $\kappa_0=\kappa_1$ for all
delays,
so the choice $\kappa_0=\kappa_1$ is always safe and free of divergences and, at
the same time, it also leads to a significant reduction of the RMSE
(provided the value of $\tau$ is not too large).

\begin{figure}
\centering
\includegraphics[width=19pc]{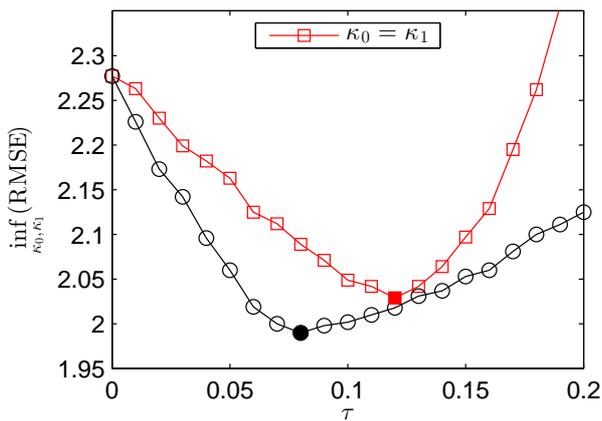}
\caption{Smallest RMSE achieved with $n_s=3$ for each value of $\tau$ (circles).
The red squares are the results after restricting to the subspace $\kappa_0=\kappa_1$ (the bisector of
the panels in Figure~\ref{ns3}).
Note that for $\tau=0$ we recover the optimal RMSE corresponding to the classical (undelayed) case.}
\label{ns3_min}
\end{figure}

To better apprehend the significance of Figure~\ref{ns3}, we represent in  
Figure~\ref{ns3_min} the infimum RMSE achieved as a function of $\tau$: 
each circle corresponds to an optimal $\tau$-dependent
combination of $\kappa_0$ and $\kappa_1$. 
For instance the smallest RMSE ($\approx1.99$), corresponding to the filled circle in Figure~\ref{ns3_min},
is obtained with $\tau_\mathrm{opt}=0.08$ for the optimal combination $(\kappa_0,\kappa_1)=(3,11.25)$,
with the delayed observation coupling strength being almost $4$ times larger
than that for the present time observation.
In fact, we find numerically that for $\tau\le 0.07$ 
the delayed coordinate greatly dominates the optimum coupling.
The trade-off option, $\kappa_0=\kappa_1$, is also displayed (squares) in 
Figure~\ref{ns3_min}. 
This simple choice substantially reduces the computational cost associated with the
tuning of 
the relaxation coefficients because the search for the infimum RMSE is restricted to
 the subspace $\kappa_0=\kappa_1$.  
With this choice we achieve 
a reasonably small RMSE ($\approx 2.04$), at $\tau=0.12$ for $\kappa_0=\kappa_1=8$,
see the filled square in Figure \ref{ns3_min}.
In summary, we numerically find $\inf(\mathrm{RMSE})=1.99$, $2.04$ and $2.28$ for the
delay-coordinate nudging with full forcing-delay optimization, 
delay optimization with the constraint 
$\kappa_0=\kappa_1$, and nudging without delay, respectively.

\begin{figure}
\centering
\includegraphics[width=18pc]{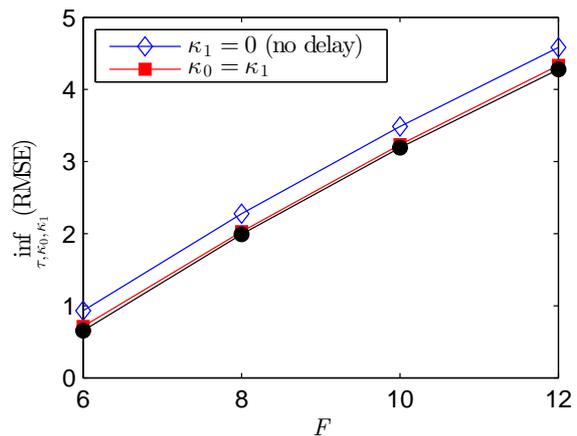}
\caption{Infimum RMSE over $\tau$, $\kappa_0$ and $\kappa_1$ 
achieved for different values of parameter $F$ (filled circles).
For comparison, the results restricting to the subspaces 
without delay (diamonds) and with equal couplings, $\kappa_0=\kappa_1$ (filled
squares)
are also represented.}
\label{f}
\end{figure}

\begin{figure*}
\centering
\includegraphics{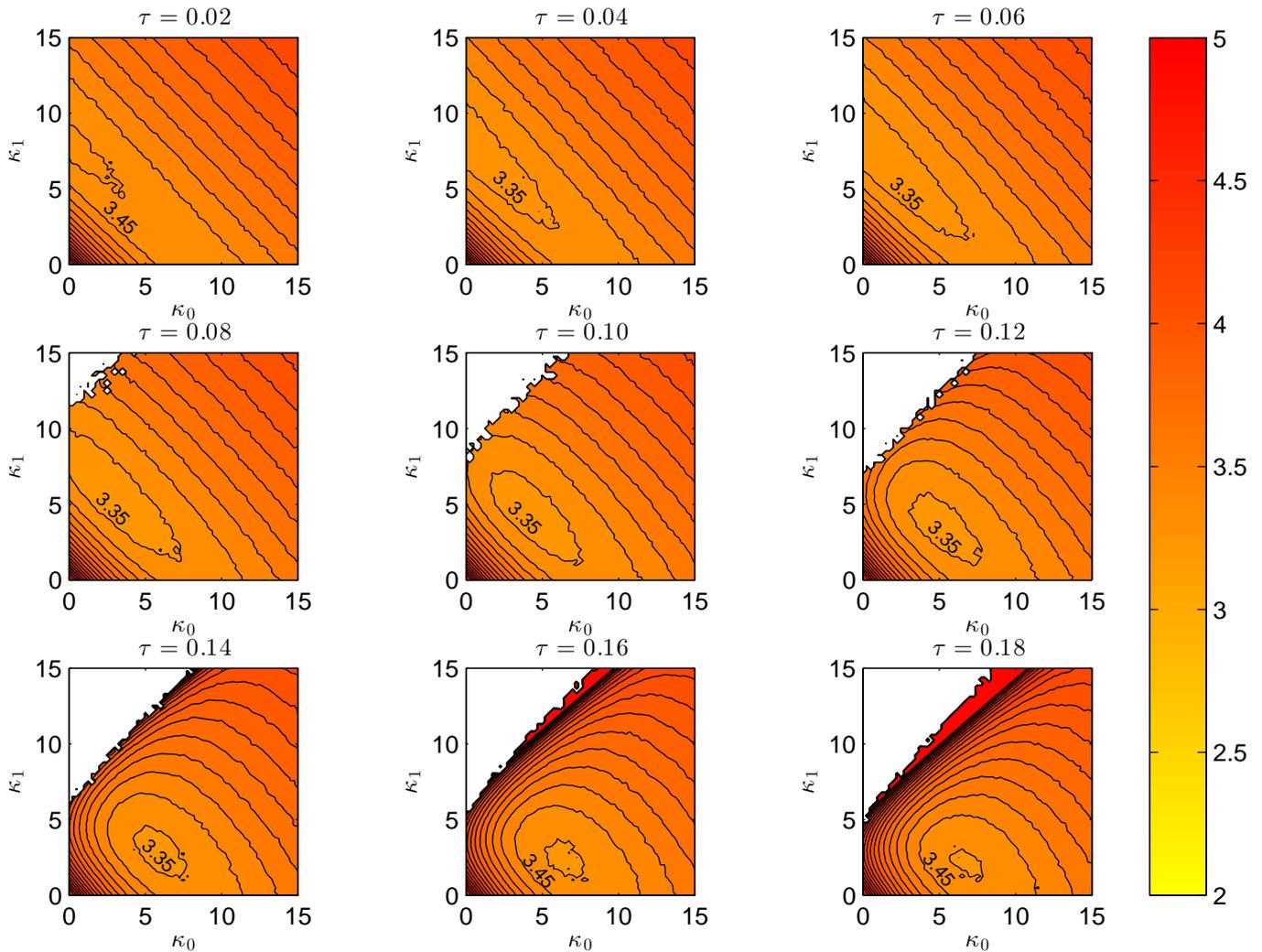}
\caption{Colour shading depicting the RMSE for $n_s=4$ and several delays.
The contour lines are at levels separated by intervals 
of $0.1$ in RMSE. 
The white region in the upper left corner of some plots 
indicates where the algorithm is found to be unstable.}
\label{ns4}
\end{figure*}

As a check for robustness of our numerical analyses, we repeated the
simulations presented in Figures~\ref{ns3} and \ref{ns3_min} for
other values of the external forcing $F$ of the model, finding no qualitative
differences in the relative performances of the algorithms.
Note that the Lorenz-96 model (\ref{l96m}) becomes progressively more chaotic
when parameter $F$ is increased:
both the number and the magnitude of the  positive Lyapunov exponents increase
monotonically with $F$.  
Figure~\ref{f} summarizes these results, where we plot with filled circles
the infimum RMSE
as $\tau$, $\kappa_0$ and $\kappa_1$ are freely varied for different values of
the external forcing $F$.
For comparison, the RMSEs obtained when we restrict ourselves to the
subspaces $\kappa_1=0$ (classical nudging without delay) 
and $\kappa_0=\kappa_1$ are also represented with blue diamonds and red squares,
respectively. We find that the results systematically improve (lower RMSE) when
using delay coordinates. 
Interestingly, the
gain with respect to the standard approach barely depends on the 
degree of `chaoticity' of the underlying dynamics, i.e.~the number of
positive Lyapunov exponents, suggesting that delay-coordinate 
nudging is systematically more effective in controlling the error growth than
the classical (nondelayed) nudging.  
Moreover the choice $\kappa_0=\kappa_1$ performs also very well, with only minor 
degradation with respect to the fully optimized case. 
This is an encouraging result as it suggests that DA can be greatly enhanced 
by restricting to the $\kappa_0=\kappa_1$ subspace,
without the need of running longer search experiments to 
determine the optimal values in the full space of couplings $\kappa_0$ and
$\kappa_1$. The optimal 
parameter values
for different values of $F$ are summarized in Table 1.

 \begin{table}
 \caption{Optimal tunning for different sets of free parameters
in the case  $n_s=3$. The left-most row lists the four different values
of $F$ used in Fig.~\ref{f}.}
\begin{tabular}{c|c|c|c}
$F$ & $\kappa_\mathrm{opt}$ (no delay)  &    $(\tau,\kappa_0=\kappa_1)_\mathrm{opt}$     &$(\tau,\kappa_0,\kappa_1)_\mathrm{opt}$ \\ \hline
6 & 11 & $(0.15,6)$ & $(0.11,2,9)$ \\ 
8 & 13 & $(0.12,8)$ &$(0.08,3,11.25)$ \\ 
10 & 15 & $(0.1,9.25)$ & $(0.06,1.75,14)$ \\ 
12 & 16 &  $(0.09,9.25)$ & $(0.05,1.25,15.75)$  
\end{tabular}\label{table1} 
\end{table}

\subsection{Case II: $n_s=4$}
\begin{figure}
\centering
\includegraphics[width=19pc]{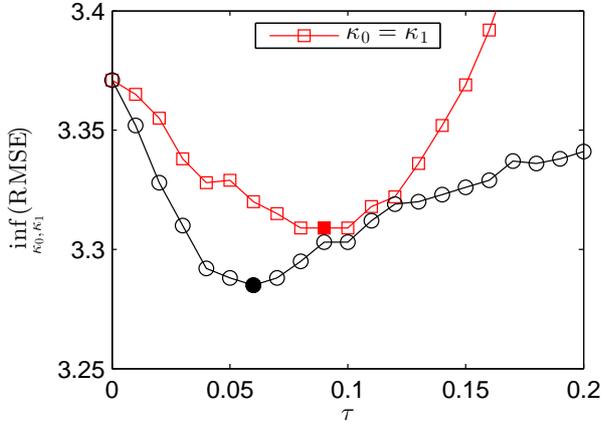}
\caption{Smallest RMSE achieved with $n_s=4$ for each value of $\tau$ (circles). 
The red squares are the results after restricting to the subspace $\kappa_0=\kappa_1$ (the bisector of
the panels in Figure~\ref{ns4}).}
\label{ns4_min}
\end{figure}

We now increase sparseness by enlarging the distance between adjacent observed
sites from $n_s=3$ to $n_s=4$,
 so that the total number of observations is now $M=15$ in a system of size
$N=60$. 
This choice makes the DA more challenging, as the state estimation  
 relies now on less observations and the RMSE will tend to grow. 
The classical nudging without delay achieves an RMSE $\approx 3.37$.  
Figure~\ref{ns4} is similar to Figure~\ref{ns3}, and qualitatively  reproduces 
the pattern found for $n_s=3$.
Delay-coordinate nudging improves systematically over the standard approach, 
with the smallest RMSE being $\approx 3.28$ for $\kappa_0 =1$, $\kappa_1 =7$
and 
$\tau=0.06$. Areas of error divergence are present in the upper left portions of the panels, 
when $\tau\ge 0.08$ and, as observed for the case $n_s=3$, the bisector $\kappa_0 = \kappa_1$ 
represents a choice for the forcing coefficients that prevents error divergence and leads to 
a minimum RMSE $\approx3.31$, lower than the standard approach. 
Note, nevertheless, that the skill gain due to the use of delay-coordinates is 
smaller for $n_s=4$ than that with $n_s=3$, see Figure~\ref{ns4_min}.
While we achieved improvements of the order of $10\%$ for $n_s=3$, the RMSE
for $n_s=4$ barely decreases a $2.5\%$, with respect to the classical nudging.

\subsection{The effect of multiple past observations: $P>2$}

\begin{figure}
\centering
\includegraphics[width=19pc]{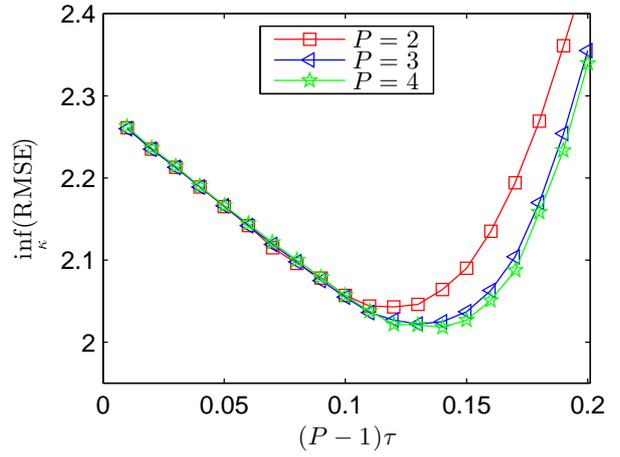}
\caption{Minimal RMSE for different values of $\tau$ and $P$. The computations
are carried out restricting to the subspace with uniform coupling
$\kappa_n=\kappa/P$, $n=0,\ldots,P-1$.}
\label{ne}
\end{figure}

We briefly investigate in this section the effect of using more than
one delayed observation, i.e.~$P>2$ in Eq.~\eqref{delayed}. 
This configuration is difficult to study systematically with numerical
experiments 
due to the large number of independent parameters that need to be tuned,
$\tau$ and
$\kappa_{n=0,\ldots,P-1}$.
As a trade-off we decided to simplify the analysis by assuming equal couplings
$\kappa_n=\kappa/P$, therefore
working with two free parameters only, namely, the delay $\tau$ and the
aggregate coupling $\kappa$.
The infimum RMSE over $\kappa$ as a function of $\tau$ is shown in Figure~\ref{ne},
for three values of $P=2,3,4$ and $n_s=3$.
Increasing from $P=2$ to $3$ leads to a clear improvement in the assimilation, 
with the smallest RMSE decreased from $2.043$ to $2.022$.
Nevertheless, a further increase to $P=4$ (i.e.~$3$ delayed
observations) does not 
yield much gain and, in fact, the improvement seems to saturate.
Interestingly, Figure~\ref{ne} indicates that the optimal total delay does not change 
significantly in the three experiments, as reflected by an almost
complete superposition of 
the curves with $(P-1)\tau$. In practice, by
increasing $P$, the optimal $\tau$ decreases so that the total delay is kept
almost constant. This finding is also consistent with the theoretical framework
in Sec.~\ref{sec_theory}.

\section{Numerical analysis: imperfect model scenario}
\label{imperf}

\begin{figure*}
\centering
\includegraphics{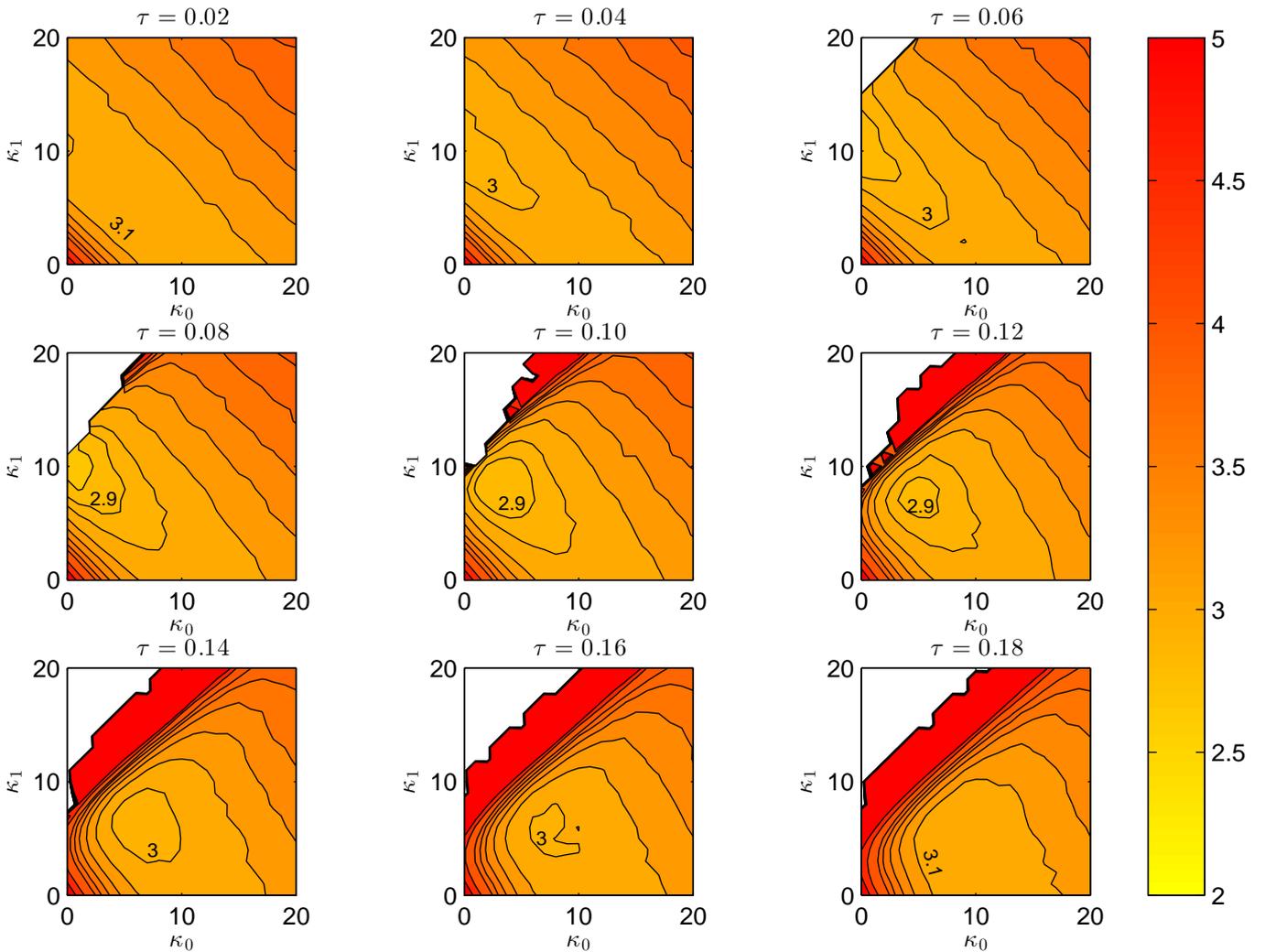}
\caption{Colour shading depicting the RMSE for $n_s=3$ and several delays. 
The model is governed by Eq.~\eqref{l96d} while the truth 
has more complex dynamics given by Eq.~\eqref{2s}.
The contour lines are at levels separated by intervals 
of $0.1$ in RMSE.}
\label{ns3_2s}
\end{figure*}

In this section, we test the robustness of the delay-coordinate nudging method
upon model imperfections.
This configuration is closer to realistic contexts where nudging is
usually applied. 
Model error can have, in practice, a number of different origins such as
parametric misspecification,
numerical discretization or the lack of some process in the description of the
reality afforded by the model. We will focus here on the latter
case. This is a particularly
demanding problem for DA (see {\it e.g.} 
\citep{MitchellCarrassi2015} and references therein). 
It is worth to mention here that 
we have also studied the parametric error case, 
with a mismatched $F$ in the model, and have
obtained equally satisfactory results (not shown).
  
To this aim, we now assume the truth dynamics includes degrees of
freedom, which are absent and not described in
the model. Specifically, we adopt the two-scale version of the
Lorenz-96 model \citep{lorenz96} for the truth:
\begin{subequations}
\label{2s}
\begin{align}
& \dot u_i=u_{i-1}(u_{i+1}-u_{i-2})-u_i+F-
\frac{h \,c}{b}\sum_{j=1}^{J} y_{j,i}\label{e.eq2a}\\
& c^{-1} \dot y_{j,i}=b\, y_{j+1,i}(y_{j-1,i}-y_{j+2,i})-y_{j,i}+
\frac{h}{b} u_i,\label{e.eq2b}
\end{align}
\end{subequations}
where the additional $y$ variables are, like $u$ variables, defined on a ring,
and thus 
$y_{J+1,i}=y_{1,i+1}$, $y_{J+2,i}=y_{2,i+1}$, $y_{0,i}=y_{J,i-1}$.
The constant $c$ sets the time scale for $y$ variables and we choose $c=10$ in
our simulations, so that $y$ evolves $10$ times faster than the
(slow) variables $u$. Accordingly, we reduced the time step of our
numerical
integration by a factor $10$, setting $dt=10^{-4}$ for both the truth,
Eq.~\eqref{2s},
and the model, Eq.~\eqref{l96d}. 
The other parameters are taken as $b=J=F=10$,
what implies that the ring of $y_{j,i}$ variables is composed of $N\times
J=600$
sites. Parameter $b$ controls the amplitude of the oscillations of $y$, 
which are ten times smaller than those of the $u$ variable. 
Moreover parameter $h$, controlling the interaction strength between $u$ and $y$ variables,
is set to $h=1$ (which means a strong interaction \citep{herrera11,pazo14});
note that $h=0$ decouples $u$ and $y$ and the perfect model scenario is
recovered. 

\begin{figure}
\centering
\includegraphics[width=19pc]{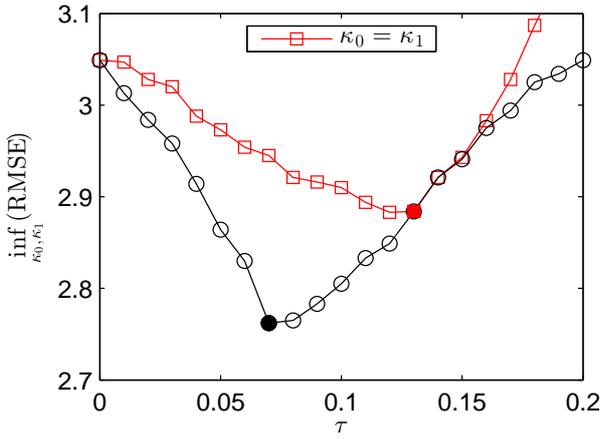}
\caption{Smallest RMSE achieved with $n_s=3$,
in the presence of model error, for each value of $\tau$ (circles). 
The red squares are the results after restricting to the subspace $\kappa_0=\kappa_1$ (the bisector of
the panels in Figure~\ref{ns3_2s}).}
\label{ns3_min2s}
\end{figure}

We select $P=2$ in the delay-coordinate nudging scheme and, in order to
lower the computational load, 
we scan the RMSE for parameters $\kappa_0$ and
$\kappa_1$ with larger step sizes, $\Delta\kappa=1$.
Results are displayed in Figures~\ref{ns3_2s} and \ref{ns3_min2s} that are the analogue of 
Figures~\ref{ns3}/\ref{ns4} and of Figures \ref{ns3_min}/\ref{ns4_min}, respectively.

The obvious consequence of model error is to make the overall
RMSE level higher than in the perfect model scenario. 
But, qualitatively, Figure~\ref{ns3_2s} clearly shows similar features than those reported
in the perfect model scenario, with the lowest RMSE area departing from 
the ordinate-axis (i.e.~increasing $\kappa_0$) for increasing $\tau$. Also
in this case
we see that the bisector, $\kappa_0=\kappa_1$, represents viable choice for the 
forcing terms ensuring better performance than in the classical nudging case
with no delay. This is 
further highlighted in Figure~\ref{ns3_min2s} that shows the improvement over
the standard 
approach by using a fully optimized delay coordinates nudging (filled circle) and 
the case of the minimum RMSE restricted to the bisector line $\kappa_0=\kappa_1$
(filled square). 
When compared with the perfect model case, Figure~\ref{ns3_min}, we see now that
the curves are much less smooth: small departures from the optimal coupling
lead to
substantially worse RMSE than the optimized choice. This higher sensitivity 
of the algorithm to coupling parameter values seems to be a direct
outcome of the presence of model error.

The results of Sections 4 and 5 are summarised in Table \ref{table2}, which 
 shows the infimum RMSE for the standard and delay-coordinate nudging.
\begin{table}
\caption{RMSE for the standard and delay-coordinate (DC in the table) nudging
with observations every 3 ($n_s=3$) or 4 ($n_s=4$) sites and in case of perfect (Perf) or imperfect model (Imp).}
\begin{tabular}{l l l l }
\toprule
\textbf{Nudging Method} & \bf{$n_s=3$} & \bf{$n_s=3$} & \bf{$n_s=4$} \\
  & \textbf{Perf} & \textbf{Imp} & \textbf{Perf} \\
\midrule
Standard  & 2.28 & 3.05 & 3.37  \\
DC (optimized) & 1.99 & 2.76 & 3.28  \\
DC (sub-optimal $\kappa_0=\kappa_1$) & 2.04 & 2.88 & 3.31 \\
\bottomrule
\end{tabular}
\label{table2}
\end{table}

\section{Mathematical discussion on
delay-coordinate nudging}
\label{sec_theory}

In this section we present 
a heuristic justification of our method for
nudging with delay coordinates and 
of its enhanced skill with respect to standard nudging.  
Our argumentation leads to an estimation of
the order of magnitude of
the optimal delay $\tau_\mathrm{opt}$ to be used in the algorithm. 

Let us begin the discussion by recalling a 
property that is well known by forecast practitioners:
in the nudging implementation one seeks to have the forcing term 
strong enough to be effective in driving the model toward the observations, but 
weak enough to avoid dynamical shocks caused by pushing the model to 
occupy states in the phase-space out of its attractor. As a general rule, 
nudging has been always implemented so that the forcing term does not exceed 
the smallest term in the prognostic equation \citep{hoke76}.    
Figure 1(b) exemplifies this effect: even with a perfect 
model ($\tilde{\mathbf g}={\mathbf g}$),
observational sparseness causes the ``synchronization manifold'', 
$\mathbf{u}=\mathbf{v}$, 
to become unstable for too large delay-coordinate couplings. And in Figure 1(c)
we see how with
even sparser observations this trade-off between weak and strong coupling
leads to a certain optimal coupling,
as complete synchronization between truth and model cannot be attained.
With this negative effect of large coupling in mind, one should seek ways of enhancing
synchronization at moderate coupling strengths. Precisely, delayed-coupling 
has been previously highlighted as a source of enhanced synchronizability in
contexts such as neural dynamics \citep{dhamala04}.

To simplify the illustration of the problem 
we consider the delay-coordinate nudging scheme in Eq.~\eqref{delayed} and assume
hereafter to have full coupling, ${\bf D}={\bf I}$, with ${\bf I}$ 
being the identity matrix in $\mathbb{R}^N$.
We expect that if stability at moderate-forced coupling 
is enhanced by delay in this case, 
then this will translate into a better assimilation
when the coupling is not dense (${\bf D}\ne {\bf I}$). This is, admittedly, 
only a plausible expectation that cannot be further formalized since we do not know
how each particular system will respond to coupling sparseness, but we expect
it to be true in the general case. Our analysis relies on the study of
infinitesimal deviations ${\mathbf{w}}$ off the synchronization manifold,
${\mathbf{v}}={\mathbf{u}}+{\mathbf{w}}$, governed by linear equations. We make
a further assumption here since assimilation error is obviously
non-infinitesimal and its dynamics nonlinear. In sum, we assume that
delay-induced gain in the synchronization stability obtained for (i)
infinitesimal error and (ii) dense coupling will yield smaller assimilation
error, when the error is not small and the coupling is not dense.

Mathematically, the dynamics of infinitesimal ${\mathbf{w}}$ with dense 
coupling, ${\bf D}={\bf I}$, 
is described by the linearisation of (\ref{delayed}):
\begin{equation}
\dot{\mathbf{w}} = \mathbf{\partial_u}\mathbf{g}({\mathbf{u}})\, {\mathbf{w}} - 
\kappa_0 \, {\mathbf{w}}-\kappa_1 \, {\mathbf{w}}^\tau
\label{linear}
\end{equation}
where $\mathbf{\partial_u}{\bf g}({\mathbf{u}})$ is the Jacobian matrix of the
dynamics evaluated at the system current state $\mathbf{u}$. 
In writing Eq.~\eqref{linear}, and hereafter, we assume $P=2$
since, as described in previous sections, 
this case already captures the effect of the delay coordinates; 
it is then trivial to extend the analysis to larger $P$ if desired.

The stability of synchronization is adequately quantified
by the exponential growth rate of $\|{\mathbf{w}}\|$, customarily referred to as
the
`transverse Lyapunov exponent' (TLE). This is defined as
the exponential growth rate of infinitesimal perturbations:
\begin{equation}
\mathrm{TLE}=\lim_{t\to\infty} \frac{1}{t} \ln\|{\mathbf{w}}\| .
\label{w}
\end{equation}
If no coupling is present
($\kappa_0=\kappa_1=0$), the chaotic dynamics of the system
elicits an exponential growth of ${\mathbf{w}}$:
$\|{\mathbf{w}}(t)\| \simeq \|{\mathbf{w}}(0)\| \exp(\mu t)$,
where $\mu$ is the Lyapunov exponent, and trivially $\mathrm{TLE}=\mu$.

\begin{figure}
\centerline{\includegraphics*[width=8cm,clip=true]{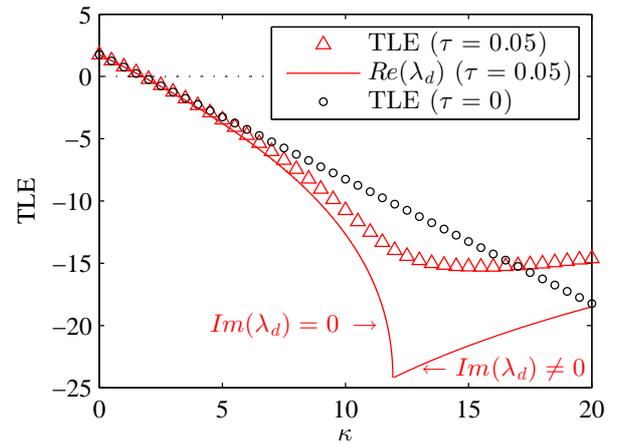}}
\caption{Transverse Lyapunov exponent as a function of $\kappa(=2 \kappa_0=2\kappa_1)$
for $\tau=0.05$ (triangles) and $\tau=0$ (circles, $\mathrm{TLE}=\mu-\kappa$). 
We observe the TLE is smaller with than without delay in the interval $3<\kappa<17$.
The solid line is $Re(\lambda_d)$, where $\lambda_d$ is the
solution of Eq.~(\ref{charac}) with the largest real part.}
\label{theory}
\end{figure}

If we now let $\kappa_0$ and $\kappa_1$ to be nonzero, it is tempting to
insert the exponential ansatz ${\mathbf{w}}(t)\sim {\mathbf{w}}(0) \exp(\lambda t)$
into (\ref{linear}), and derive the characteristic equation:
\begin{equation}
\lambda=\mu-\kappa_0-\kappa_1 e^{-\lambda\tau}
\label{charac}
\end{equation}
Unfortunately, this equation does not exactly describe the asymptotic dynamics of $\mathbf{w}$,
unless $\mathbf{\partial_u g}$ is constant (like occurs if $\bf u$ is
a fixed point), due to the subtle interplay between a fluctuating matrix and the delay. 
Nonetheless, among the spectrum of solutions of Eq.~(\ref{charac}),
the branch that correspond to the largest real part, denoted here by
$\lambda_d$,
yields an estimation of the TLE.
The estimation $Re(\lambda_d)\approx\mathrm{TLE}$
is exact for $\tau=0$ and progressively deteriorates as $\tau$ increases.
In Figure~\ref{theory} we illustrate with an example 
(setting $\tau=0.05$ and $\kappa_0=\kappa_1=\kappa/2$)
how the characteristic Eq.~\eqref{charac} estimates the TLE.
In the figure we represent with triangles
the exact TLE numerically obtained via integration of Eq.~\eqref{linear},
and $Re(\lambda_d)$ ---easily obtained after inserting the Lyapunov exponent $\mu\simeq 1.75$
into (\ref{charac})--- is depicted by a solid line. 
In spite of the evident roughness of the approximation,
the crucial point of our reasoning is that for both the true TLE and the approximation $Re(\lambda_d)$
there is a range of $\kappa$ values where
the TLE is smaller with the use of delay-coordinates than without,
$\mathrm{TLE}(\tau=0)=\lambda_d=\mu-\kappa$.
In other words, delay favours the stability of synchronization with respect
to the delay-free case; compare the triangles and the small circles in Figure~\ref{theory}.

In view of the cusp in the theoretical continuous line in Figure~\ref{theory}, 
it is useful to search for a relationship between
$\tau$ and the ``optimal'' value $\kappa_*$ where the minimum of $Re(\lambda_d)$ is located.
This relationship is given by the transcendental equation:
$\tau=(2/\kappa_*)\exp[(\mu-\kappa_*/2)\tau-1]$.
This permits an estimation of the order of magnitude of the optimal delay.
In our example with the Lorenz-96 model, given the 
wide range where the optimal coupling is reasonably expected to be, say $\kappa_*\in[2,20]$,
the transcendental equation predicts $\tau\in[0.029,0.56]$. 
This is only an estimation of the order of magnitude of
$\tau$, but can be a valuable information as a first guess.
Note that the optimal $\tau$ are $0.12$ ($n_s=3$) and $0.09$ ($n_s=4$)
in Figures~\ref{ns3_min} and \ref{ns4_min}, respectively.

As a final comment, we resort to Eq.~\eqref{charac} to rationalize why choosing 
$\kappa_1>\kappa_0$ may destabilize the nudging algorithm, particularly for non-small delays (the
white regions in Figures~\ref{ns3}, \ref{ns4}, and \ref{ns3_2s}).
Note first that for non-small $\tau$, and unless $\kappa$ is too small,
the leading solution, i.e.~the one with largest real part, has
$Im({\lambda_d})\ne 0$ (this branch is, for instance, the solid line
at the right of the cusp point in Figure~\ref{theory}).
Given our expectation of dominant imaginary eigenvalues in Eq.~\eqref{charac},
we may foresee that oscillating modes (with nonzero imaginary part) dominate the fluctuations
along the chaotic trajectory when delay is large. The key point is that 
those modes are associated with certain finite-time exponents $\Lambda=\Lambda_r+i\Lambda_i$,
which leads to a coefficient $e^{-\Lambda_r \tau} e^{-i \Lambda_i \tau}$ in $\kappa_1$,
see Eqs.~\eqref{linear} and \eqref{charac}.
The source of instability is the factor $e^{-i \Lambda_i \tau}$. In the worst
situation ($\Lambda_i \tau=\pm \pi,\pm3\pi,\ldots$) the sign of the $\kappa_1$ 
term is flipped, 
so it becomes a destabilizing contribution. This 
line of reasoning also suggests that if $\kappa_0\ge\kappa_1$ the non-delayed part
preserves the stability of the nudging algorithm, while choosing $\kappa_0<\kappa_1$
can not guarantee the stability of the nudging algorithm, specially if the delay
is non-small or $\kappa_1\gg\kappa_0$.
This argument is mainly intuitive and heuristic. 
As such it does not represent a mathematical proof but rather 
 a reasoning of what contributions are expected to affect the stability of the algorithm 
and guiding the practical implementation of the delay-coordinated nudging.

\section{Conclusions}

A new approach for DA, the delay-coordinate nudging, has
been proposed and extensively investigated. It merges the use of delay
coordinates,
typical of phase-space reconstruction by temporal embedding,
with the standard nudging technique. In the delay-coordinate nudging, both
present and past observations are used
to nudge the model evolution at each time-step, with negligible
increase in computational cost.

Delay-coordinate nudging has been numerically compared with standard nudging
in the context of the Lorenz-96 model, for different observational densities,
and in the case of both perfect and imperfect model. In all the circumstances
considered
the new approach permits to achieve a 
better skill, measured using the RMSE, than that obtained with standard
nudging. The number of parameters to be optimized--- namely, the coupling
coefficients---
grows with the number of delay coordinates. However numerical results suggest that:
{\it (i)} using the same coefficients for all observations already provides a
systematic improvement
over the classical nudging (only slightly worse than the fully optimized case),
and  
{\it (ii)} the performance improvement almost saturates when the number of 
used delayed observations is larger than $3$.
Both facts crucially reduce the optimization cost and 
open the path for the implementation of delay-coordinate nudging in realistic
frameworks. 

A connection between the optimal length of the delay, $\tau_\mathrm{opt}$,
and the dominant Lyapunov exponent of the model dynamics is 
put forward
by studying the stabilization conditions of the synchronization manifold.
In the case of using only one delayed observation ($P=2$), the estimate 
of $\tau_\mathrm{opt}$ obtained  
is consistent with the
numerical observations. 
The same line of reasoning has allowed us to interpret the existence
of an unstable regime for the delay-coordinate        
nudging scheme, observed numerically for certain couplings satisfying
$\kappa_1>\kappa_0$, 
as arising from oscillating instabilities along the chaotic trajectory. 
While a formal proof of this relation has not been established, heuristic 
arguments were provided to support it. 

Summarizing, delay-coordinate nudging represents a straightforward amelioration 
over the standard approach with a small increase of the computational cost.
As a continuous data assimilation method, the present approach is also 
related to recent studies where incomplete data are continuously 
assimilated to retrieve the signal in fluido-dynamical systems,
and that provide an analysis of the temporal and spatial resolution of the observations 
to achieve a satisfactorily reconstruction of the system's state 
\citep{Blomker_et_al_2013, Farath_et_al_2015}. 
{\color{black} Indeed in these latter works a continuous time 3DVar is adopted implying the
use of a spatial extended error covariance matrix so allowing for the propagation of 
the data forcing to unobserved areas. The extension of the present delay-coordinate formulation 
in combination with a continuos time 3DVar represents an interesting line of follow-on research.} 
Further improvements, inspired in advanced attractor reconstruction methods
for systems with multiple time scales \citep{pecora07}, 
may consist in adopting delayed coordinates unevenly distributed in time.
Another relevant direction of research is the combination of delay-coordinate
nudging with advanced initialisation methods for coupled models
that account for model bias.
All in all, the application to more realistic, observational and model, scenarios 
is a natural direction of prosecution of this study.

\acks

We thank Miguel A. Rodr\'iguez for fruitful and interesting discussions.
We acknowledge support by MINECO (Spain) under project No.~FIS2014-59462-P.
DP acknowledges support by MINECO (Spain) under the Ram\'on y Cajal programme.
AC was supported by the EU-FP7 project SANGOMA under grant contract 283580.


\end{document}